\documentclass{article}

\title{Conformal Transformation of the Schr\"{o}dinger Equation for Central Potential Problems in Three-Dimensions}
\author{Robert Ducharme}

\begin{document}
\maketitle

\centerline{151 Fairhills Dr., Ypsilanti, MI 48197}
\centerline{E-mail: ducharme01@comcast.net}

\begin{abstract}
In a recent paper, it has been shown the Schr\"{o}dinger equation for the three-dimensional harmonic oscillator can be simplified through the use of an isometric conformal transformation. Here, it is demonstrated that the same transformation technique is also applicable to the Schr\"{o}dinger equation for the hydrogen atom. This approach has two interesting features. Firstly, it eliminates potential fields from the Schr\"{o}dinger equation. The Coulomb and harmonic binding terms are instead represented as imaginary parts of complex time. Secondly, the method leads to a general relationship between potential energy and ground state energy that encompasses both the hydrogen atom and the harmonic oscillator as special cases. 
\end{abstract}

\section{Introduction}
Conformal mapping \cite{ZN} is a heavily studied coordinate transformation technique that has found numerous practical applications in many science and engineering disciplines. Most of these applications are for two-dimensional problems but the scope of the technique is not limited to two-dimensions. Liouville's theorem \cite{DEB} in fact shows that higher dimensional conformal maps are possible but must be composed of translations, similarities, orthogonal transformations and inversions. In a recent paper \cite{RJD}, an isometric conformal transformation has been used to simplify the Schr\"{o}dinger equation for the three-dimensional harmonic oscillator. The purpose of this paper is to show the same transformation technique can also be used to simplify the Schr\"{o}dinger equation of the hydrogen atom.

An hydrogen atom consists of a single electron of energy E  with a spatial displacement $x_i$ $(i=1,2,3)$ from the center of mass of the atom. The source of the Coulomb field binding the electron to the atom is the much heavier proton. Both the electron and the proton may be assumed to exist at the same time t. The goal of section 2 of this paper is to introduce an isometric conformal mapping of the form $z_i= x_i,s=t+\imath f(|x_i|,E)$ where $f$ is a real function and $\imath=\sqrt{-1}$. It is clear from inspection that this passive transformation does no more than introduce an imaginary shift in the time of the two related coordinate systems.

It is convenient to write the complex conjugate form of the $(z_i,s)$-coordinates as $(z_i^*,s^*)$ even though $z_i^*=z_i$ since $z_i^*$ and $z_i$ still belong to different coordinate systems. This distinction is shown to be most evident in the computation of the partial derivatives $\partial / \partial z_i$ and $\partial / \partial z_i^*$ from the chain rule of partial differentiation as these evaluate differently in their conjugate coordinate systems. One further topic to be introduced in section 2 is the Cauchy-Riemann equations that are necessary to determine if a function transformed into the $(z_i,s)$-coordinate system has well defined partial derivatives. 

The Schr\"{o}dinger equation for the hydrogen atom is presented in section 3 alongside a complete set of eigensolutions. It is shown that both these results simplify in terms of $(z_i,s)$-coordinates and that the eigensolutions are holomorphic. This argument is used to demonstrate that a Coulomb term may be introduced into the free-field Schr\"{o}dinger equation by adding an imaginary component to the world time of the particle. The imaginary part of time has a binding effect on the particle equivalent to including a Coulomb potential in the Hamiltonian.

In section 4, the results of applying a similar conformal transformation technique to both the harmonic oscillator and the hydrogen atom are compared. It is shown, firstly, the transformed Schr\"{o}dinger equation is identical in form for both problems. Secondly, there is no need for potential fields in the complex $(z_i,s)$-coordinate system since the binding terms are included as the imaginary parts of the complex time of the particles. Thirdly, the method generates a novel relationship between potential energy and ground state energy that encompasses both the hydrogen atom and harmonic oscillator as special cases.

\section{Conformal Mapping}
The task ahead is to present a isometric conformal transformation relating a real $(x_i,t)$-coordinate system and a complex $(z_i,s)$ coordinate system. This mapping is to be applied in section 3 to a hydrogen atom consisting of an electron of mass $m_e$ and total energy E in the electrostatic field of a proton of mass $m_p$. It is convenient to express it in the form
\begin{equation} \label{eq: conftrans1}
z_{i} = x_{i}, \quad s = t - \imath \frac{\hbar}{E}\frac{r}{\alpha_0} 
\end{equation}
where $r=|x_i|$,
\begin{equation} \label{eq: bohrRadius}
\alpha_0 = 	\frac{4 \pi \epsilon \hbar^2}{\mu e^2}
\end{equation}
is the Bohr radius, -e is the charge on the electron, $\epsilon_0$ is the permittivity of free space, $\hbar$ is Planck's constant divided by $2\pi$ and $\mu = m_e m_p /(m_e+m_p)$ is the reduced mass of the electron. In the $(x_i,t)$-coordinate system, the electron has a spatial displacement $x_i$ from the proton but shares the same world time t. Similarly, in the $(z_i,s)$-coordinate system, the electron has a displacement $z_i = x_i$ from the proton but shares the same complex time s. The isometric nature of the transformation therefore follows from the result $|z_i|=|x_i|$. It is also clear that the complex time $s$ is translated through an imaginary displacement from the real time $t$. 

In the application of complex coordinates to express physical problems, there is generally going to be both a complex and a complex conjugate coordinate representation for each individual problem. In the present case, the complex conjugate of eq. (\ref{eq: conftrans1}) is
\begin{equation} \label{eq: conftrans2}
z_{i}^* = x_{i}, \quad s^*= t + \imath \frac{\hbar}{E}\frac{r}{\alpha_0} 
\end{equation}
Naturally, there must also be inverse transformations mapping the complex and complex conjugate representations of the problem back into a single physical coordinate system. The inverses of the transformations (\ref{eq: conftrans1}) and (\ref{eq: conftrans2}) are readily shown to be
\begin{equation} \label{eq: inv_ict1}
x_{i} = z_{i}, \quad t = s + \imath \frac{\hbar}{E}\frac{r_z}{\alpha_0} 
\end{equation}
\begin{equation} \label{eq: inv_ict2}
x_{i} = z_{i}^*, \quad t = s^* - \imath \frac{\hbar}{E}\frac{r_z}{\alpha_0}  
\end{equation}
$(r_z =|z_i|)$ respectively.  

It is now interesting to investigate properties of derivatives with respect to complex 4-position coordinates. In particular, the chain rule of partial differentiation gives
\begin{equation}  \label{eq: complexDiff1}
\frac{\partial}{\partial s}  
= \frac{\partial t}{\partial s} \frac{\partial}{\partial t}
+ \frac{\partial x_{i}}{\partial s} \frac{\partial}{\partial x_{i}}
= \frac{\partial}{\partial t} 
\end{equation}
\begin{equation} \label{eq: complexDiff2}
\frac{\partial}{\partial s^*}  
= \frac{\partial t}{\partial s^*} \frac{\partial}{\partial t}
+ \frac{\partial x_{i}}{\partial s^*} \frac{\partial}{\partial x_{i}}
= \frac{\partial}{\partial t} 
\end{equation}
\begin{equation} \label{eq: complexDiff3}
\frac{\partial}{\partial z_i} 
= \frac{\partial x_i}{\partial z_i} \frac{\partial}{\partial x_i}
+ \frac{\partial t}{\partial z_i} \frac{\partial}{\partial t}
= \frac{\partial}{\partial x_i} + \frac{\imath}{\alpha_0} \frac{x_i}{r}   \frac{\hbar}{E}  \frac{\partial}{\partial t}
\end{equation}
\begin{equation} \label{eq: complexDiff4}
\frac{\partial}{\partial z_i^*} 
= \frac{\partial x_i}{\partial z_i^*} \frac{\partial}{\partial x_i}
+ \frac{\partial t}{\partial z_i^*} \frac{\partial}{\partial t}
= \frac{\partial}{\partial x_i} - \frac{\imath}{\alpha_0} \frac{x_i}{r}   \frac{\hbar}{E}  \frac{\partial}{\partial t}
\end{equation}
Note, eqs. (\ref{eq: complexDiff1}) and (\ref{eq: complexDiff3}) have been obtained using eq. (\ref{eq: inv_ict1});  eqs. (\ref{eq: complexDiff2}) and (\ref{eq: complexDiff4}) are based on eq. (\ref{eq: inv_ict2}). It has also been assumed in deriving eqs. (\ref{eq: complexDiff1}) through (\ref{eq: complexDiff4}) that
\begin{equation} \label{eq: complexDiff5}
\frac{\partial z_i}{\partial s}=\frac{\partial s}{\partial z_i} = \frac{\partial z_i^*}{\partial s^*}=\frac{\partial s^*}{\partial z_i^*}=0
\end{equation}
indicating that the coordinates $z_i$ and $s$ are independent of each other as are the complex conjugate coordinates $z_i^*$ and $s^*$. This assumption is readily validated using eqs. (\ref{eq: complexDiff1}) through (\ref{eq: complexDiff4}) to directly evaluate each of the derivatives in eq. (\ref{eq: complexDiff5}) in $(x_i,t)$-coordinates. 

In further consideration of eqs. (\ref{eq: conftrans1}), it is convenient to write $s=t+iy$ where $y = -(\hbar / \alpha_0 E)r$. The requirement for a continuously differentiable function $\tau(s)=g(t,y)+ih(t,y)$ to be holomorphic is then for the real functions g and h to satisfy the set of Cauchy-Riemann equations
\begin{equation}
\frac{\partial g}{\partial y}=\frac{\partial h}{\partial t} \quad
\frac{\partial h}{\partial y}=-\frac{\partial g}{\partial t} 
\end{equation}
or equivalently
\begin{equation} \label{eq: creq}
\frac{\partial^2 \tau}{\partial t^2} + \frac{\alpha_0^2 E^2}{\hbar^2} \frac{\partial^2 \tau}{\partial r^2} = 0
\end{equation}
It is thus concluded that a function $\psi(x_i,t)$ will also have an equivalent holomorphic form $\theta(z_i)\tau(s)$ in the complex $(z_\mu, s)$-coordinate system providing it is separable and $\tau$ satisfies eq. (\ref{eq: creq}). Here, it is understood that the domain of the Cauchy-Riemann equations in this problem is the complex plane containing s. The Cauchy-Riemann equations put no restriction at all on the form of the function $\theta(z_i)$ since $z_i$ and $s$ are independent coordinates and $z_i$ belongs to a real three-dimensional space.

\section{The Hydrogen Atom}
The Schr\"{o}dinger equation determining the wavefunction $\psi(x_i, t)$ for an electron bound in the Coulomb field of a proton can be expressed in the form
\begin{equation} \label{eq: schrod1}
-\frac{\hbar^2}{2 \mu} \nabla^2 \psi - \frac{e^2}{4 \pi \epsilon_0 r}\psi = E\psi
\end{equation}
where $\nabla^2 ( = \partial^2 / \partial x_i^2)$ is the Laplacian operator and
\begin{equation} \label{eq: schrod2}
E\psi = \imath \hbar \frac{\partial \psi}{\partial t} 
\end{equation}
gives the total energy of the electron.

The solution \cite{DFL} to eqs. (\ref{eq: schrod1}) and (\ref{eq: schrod2}) is most convenient to express in spherical polar coordinates $(r,\theta,\phi)$. In this case, it takes the separable form
\begin{eqnarray} \label{eq: psi1} 
\psi(r,\theta,\phi,t) = R(r)Y_{lk}(\theta, \phi)\exp(-\imath Et / \hbar)
\end{eqnarray}
where 
\begin{eqnarray} \label{eq: phi1} 
R_{nl}(r) =  C_{nl} \exp \left(-\frac{r}{n\alpha_0} \right)\left( \frac{2r}{n\alpha_0}\right)^l L_{n-l-1}^{2l+1}\left(\frac{2r}{n\alpha_0}\right)
\end{eqnarray}
$L_{n-l-1}^{2l+1}$ are associated
Laguerre polynomials and $n,l,k$ are the hydrogenic quantum numbers. The normalized angular component $Y_{lk}(\theta, \phi)$ is included here for completeness but is unaffected by the conformal transformation. The normalization constant for the radial component is
\begin{equation}
C_{nl} = \left( \frac{e^2 \mu}{\hbar^2} \right)^{3/2} \frac{2}{n^2} \sqrt{\left( \frac{(n-l-1)!}{[(n+l)!]^3} \right)}
\end{equation}

In developing the connection between complex $(z_i,s)$-coordinates and the hydrogen atom, eqs. (\ref{eq: complexDiff3}), (\ref{eq: complexDiff4}) and (\ref{eq: schrod2}) can be combined to give
\begin{equation} \label{eq: complexDiff6}
\frac{\partial}{\partial z_i} 
= \frac{\partial}{\partial x_i} + \frac{x_i}{\alpha_0r}  
\end{equation}
\begin{equation} \label{eq: complexDiff7}
\frac{\partial}{\partial z_i^*} 
= \frac{\partial}{\partial x_i} - \frac{x_i}{\alpha_0r}
\end{equation}
These results lead to the operator relationship
\begin{equation}\label{eq: qprop1}
 \frac{\hbar^2}{2\mu}\frac{\partial}{\partial z_i^* \partial z_i} + \frac{\hbar^2}{2\mu \alpha_0^2}  = \frac{\hbar^2}{2\mu}\frac{\partial}{\partial x_i^2}+ \frac{e^2}{4\pi \epsilon_0 r} 
\end{equation}
enabling the Schr\"{o}dinger equation (\ref{eq: schrod1}) for the harmonic oscillator to be expressed in the concise form
\begin{equation}\label{eq: complexSchrod1}
-\frac{\hbar^2}{2\mu}\frac{\partial \psi}{\partial z_i^* \partial z_i} - \frac{\hbar^2}{2\mu \alpha_0^2}  \psi= E \psi
\end{equation}
It is also readily shown using eqs. (\ref{eq: complexDiff1}) and (\ref{eq: schrod2}) that
\begin{equation} \label{eq: complexSchrod2}
E\psi = \imath \frac{\partial \psi}{\partial s} 
\end{equation}
Eqs. (\ref{eq: complexSchrod1}) and (\ref{eq: complexSchrod2}) together, therefore, constitute a complete description of the hydrogen  atom in terms of $(z_i,s)$-coordinates. On comparing eq. (\ref{eq: schrod1}) and (\ref{eq: complexSchrod1}), it is clear that the Coulomb potential term in the original Schr\"{o}dinger equation is replaced by a constant term in complex coordinates.

The oscillator function (\ref{eq: psi1}) is readily transformed into the spherical polar form of $(z_i,s)$-coordinates using eqs. (\ref{eq: conftrans1}) to give
\begin{eqnarray} \label{eq: psi3} 
\psi(r_z,\theta_z,\phi_z,s) = R(r_z)Y(\theta_z, \phi_z)\tau(s)
\end{eqnarray}
where
\begin{eqnarray} \label{eq: Rz} 
R_{nl}(r_z) =  C_{nl} \exp \left(\frac{r_z}{\alpha_0}-\frac{r_z}{n\alpha_0} \right)\left( \frac{2r_z}{n\alpha_0}\right)^l L_{n-l-1}^{2l+1}\left(\frac{2r_z}{n\alpha_0}\right)
\end{eqnarray}
\begin{eqnarray} \label{eq: psi4} 
\tau(s) = \exp(-\imath Es / \hbar)
\end{eqnarray}
It is notable that eq. (\ref{eq: psi1}) and (\ref{eq: psi3}) are similar except for an additional $\exp(r/\alpha_0)$ factor in eq. (\ref{eq: Rz}). It is also notable that $\tau(s)$ is a continuously differentiable solution of eq.(\ref{eq: creq}) thus demonstrating that the oscillator function $\psi(z_i, s)$ is holomorphic.

In consideration of the foregoing arguments, it is of interest that eqs. (\ref{eq: conftrans1}) reduces to the form $z_i=x_i, s=t$ on setting $e=0$. It also apparant that eq. (\ref{eq: complexSchrod1}) reduces to the free field form of the Schr\"{o}dinger equation under these same conditions. The converse of this argument is that the Coulomb binding term may be introduced into the free-field Schr\"{o}dinger equation through the replacement $t \rightarrow t - \imath \frac{ \hbar r}{E \alpha_0}x^2$ exactly equivalent to the more usual approach of adding the Coulomb potential into the Hamiltonian for the electron.

\section{Comparison to the Harmonic Oscillator}
A similar conformal transformation technique has been applied to the hydrogen atom in this paper and to the three-dimensional harmonic oscillator in a previous one \cite{RJD}. The general approach in both cases is to use a mapping of the form
\begin{equation} \label{eq: gentrans}
z_{i} = x_{i}, \quad s = t - \imath \frac{\hbar}{E}\left(\frac{r}{b}\right)^{\lambda} 
\end{equation}
to eliminate the central potential term $V$ from the Schr\"{o}dinger equation 
\begin{equation} \label{eq: schrod3}
-\frac{\hbar^2}{2 \mu} \frac{\partial^2 \psi}{\partial x_i^2} + V\psi = E\psi, \quad i\hbar\frac{\partial \psi}{\partial t} = E \psi
\end{equation}
giving
\begin{equation} \label{eq: schrod4}
-\frac{\hbar^2}{2 \mu} \frac{\partial^2 \psi}{\partial z_i^* \partial z_i} = (E-E_0)\psi, \quad i\hbar\frac{\partial \psi}{\partial s} = E \psi
\end{equation}
Here, the hydrogen atom and the harmonic oscillator differ in the index $\lambda$, the length scale $b$, the ground state energy $E_0$ and the potential V. The comparison between the two cases is made in table \ref{tab:comparison}. 

\begin{table}[h]
\centering

\begin{tabular}{|c|c|c|c|c|}
\hline
\parbox[c][0.3in][c]{0.75in}{System} & $\lambda$ & $b$ & $V$ & $E_0$ \\ \hline
\parbox[c][0.4in][c]{0.75in}{hydrogen atom} & 1 & $\frac{4 \pi \epsilon_0 \hbar^2}{\mu e^2}$ & $\frac{-e^2}{4 \pi \epsilon_0 r}$ & $\frac{-e^4 \mu}{32 \pi^2 \epsilon_0^2 \hbar^2}$ \\ \hline
\parbox[c][0.4in][c]{0.75in}{harmonic oscillator} & 2 & $\sqrt{\frac{2\hbar}{\mu\omega}}$ & $\frac{1}{2} \mu \omega^2 r^2$ & $\frac{3}{2} \hbar \omega$ \\ \hline
\end{tabular}
\caption{Comparison of basic parameters for the hydrogen atom and harmonic oscillator.}
\label{tab:comparison}
\end{table}

The differential form of eq. (\ref{eq: gentrans}) is readily obtained using the chain rule for partial differentiation. This gives
\begin{equation}  \label{eq: complexDiff10}
\frac{\partial}{\partial s}  
= \frac{\partial}{\partial s^*}
= \frac{\partial}{\partial t}
\end{equation}
\begin{equation} \label{eq: complexDiff11}
\frac{\partial}{\partial z_i} 
= \frac{\partial}{\partial x_i} + b^{-\lambda} \frac{\partial r^\lambda}{\partial x_i}  
\end{equation}
\begin{equation} \label{eq: complexDiff12}
\frac{\partial}{\partial z_i^*} 
= \frac{\partial}{\partial x_i} - b^{-\lambda} \frac{\partial r^\lambda}{\partial x_i}  
\end{equation}
having made use of eq. (\ref{eq: schrod2}). It follows from these results that the equivalence of eqs. (\ref{eq: schrod3}) and (\ref{eq: schrod4}) depends on the condition
\begin{equation} \label{eq: EV-relation}
\frac{1}{b^{2\lambda}} \left(\frac{\partial r^\lambda}{\partial x_i}\right)^2 
- \frac{1}{b^\lambda} \frac{\partial^2 r^\lambda}{\partial x_i^2}  
= \frac{2 \mu}{\hbar^2}(V-E_0)
\end{equation}
that is readily solved to yield
\begin{equation}
V=\frac{-\hbar^2}{\mu br}, \quad E_0=\frac{-\hbar^2}{2 \mu b^2}
\end{equation}
for the hydrogen atom ($\lambda=1$) and
\begin{equation}
V=\frac{2\hbar^2r^2}{\mu b^4}, \quad E_0=\frac{3\hbar^2}{\mu b^2}
\end{equation}
for the harmonic oscillator ($\lambda=2$). 

For the harmonic oscillator, partial derivatives in $z_i$-space can be scaled to give the lowering $\hat{a}_\mu$ and raising $\hat{a}_\mu^\dag$ operators  
\begin{equation}  \label{ladder_nr}
\hat{a}_i = \frac{b}{2}\frac{\partial}{\partial z_i}, \quad \hat{a}_i^\dag = -\frac{b}{2}\frac{\partial}{\partial z_i^*}
\end{equation} 
By comparison the partial derivatives in the $z_i$-space of the hydrogen atom do not appear to have such a straightforward interpretation. Although it is interesting the ground state wavefunctions for both the harmonic oscillator and the hydrogen atom can be calculated from the conditions
\begin{equation} 
\frac{\partial \psi_0}{\partial z_i} = 0, \quad \imath \hbar \frac{\partial \psi_0}{\partial s} = E \psi_0
\end{equation} 
equivalent to the familiar expressions $\hat{a}_\mu\psi_0=0$ and $\imath \hbar \partial \psi_0 / \partial t = E \psi_0$ in the case of the harmonic oscillator. The solution to these equations in $(z_i,s)$-coordinates is $\exp(-\imath Es / \hbar)$.

\section{Concluding Remarks}
It has been shown the concept of a potential field can be eliminated from the mathematical description of both the non-relativistic quantum harmonic oscillator and the hydrogen atom through the use of an isometric conformal transformation. In the transformed coordinate system, time is a complex quantity. The real part of this complex time is the world time; the imaginary part is responsible for binding the particles into their respective systems. The method has been applied here to derive a novel relationship between potential energy and ground energy that includes the hydrogen atom and harmonic oscillator as special cases.

\end{document}